# Trion induced negative photoconductivity in monolayer MoS$_2$

C. H. Lui[1], A. J. Frenzel[1,2], D. V. Pilon[1], Y.-H. Lee[3,4], X. Ling[3], G. M. Akselrod[1], J. Kong[3], and N. Gedik[1*]

[1]*Department of Physics, Massachusetts Institute of Technology, Cambridge, Massachusetts 02139, USA*
[2]*Department of Physics, Harvard University, Cambridge, Massachusetts 02138, USA*
[3]*Department of Electrical Engineering and Computer Science, Massachusetts Institute of Technology, Cambridge, Massachusetts 02139, USA*
[4]*Department of Materials Science and Engineering, National Tsing Hua University, Hsinchu 30013, Taiwan*
*email: gedik@mit.edu

**Abstract:** Optical excitation typically enhances electrical conduction and low-frequency radiation absorption in semiconductors. We have, however, observed a pronounced transient *decrease* of conductivity in doped monolayer molybdenum disulfide (MoS$_2$), a two-dimensional (2D) semiconductor, under femtosecond laser excitation. In particular, the conductivity is reduced dramatically down to only 30% of its equilibrium value with high pump fluence. This anomalous phenomenon arises from the strong many-body interactions in the system, where photoexcited electron-hole pairs join the doping-induced charges to form trions, bound states of two electrons and one hole. The resultant increase of the carrier effective mass substantially diminishes the carrier conductivity.

The influence of electron-electron correlations on the physical properties of materials is a topic of considerable fundamental and practical importance for nanoscience [1]. Atomically thin transition-metal dichalcogenides (TMDs, e.g. MoS$_2$, MoSe$_2$, WS$_2$, WSe$_2$), a new class of two-dimensional (2D) semiconducting materials beyond graphene, provide an excellent platform for investigating such phenomena [2, 3]. These materials exhibit remarkable characteristics, such as strong photoluminescence (PL) [4, 5], excellent (opto)electronic performance [6, 7] and controllable valley polarization [8-12]. Due to strong quantum confinement and reduced screening in the strict 2D limit, charge carriers in monolayer TMDs are subject to substantial Coulomb interactions. These interactions cause photoexcited electron-hole pairs to form tightly bound excitons [13-23], which can further capture additional charges to form trions (charged excitons) in samples with excess charges. Recent experiments have shown that trions in monolayer TMDs possess unusually high dissociation energies (20 - 50 meV, the energy needed to break a trion into an exciton and a free charge), and their concentration and spin configuration can be controlled efficiently by electrical gate and optical helicity [12, 16, 24, 25]. Although these strong many-body effects have been observed in 2D TMD systems, their influence on the materials' intrinsic conductive behavior and implication for device applications have not been explored thus far.

In this letter, we report the first experimental signature of profound trionic effect on the conductive properties of atomically thin TMD materials. By using time-resolved terahertz (THz) spectroscopy [26, 27], we observed an anomalous transient *decrease* of THz conductivity in doped monolayer MoS$_2$ after femtosecond laser excitation at temperatures T = 4 - 350 K. In particular, the conductivity of monolayer MoS$_2$ is substantially reduced to ~30% of its equilibrium value at high pump fluence. This behavior contrasts sharply with those found in multilayer and bulk MoS$_2$, as well as in other conventional semiconductors, such as Si, Ge and GaAs, where photoexcited carriers increase the materials' conductivity [26, 27]. Our results reflect the strong trionic effect in monolayer MoS$_2$, where photoexcited electron-hole pairs are bound to the doping-induced free charges and form trions within a few picoseconds. Instead of promoting the free-carrier population, interband photoexcitation substantially increases the effective mass of charge carriers, and consequently diminishes their mobility and conductivity.

To reveal the intrinsic conductive properties of monolayer MoS$_2$, we used time-domain THz spectroscopy. This approach avoids the complication of electrical contacts in transport experiments, and when applied with femtosecond laser excitation, it can probe short-lived excited carriers prior to trapping or recombination [26, 27]. Our experiment employed a 5-kHz Ti:sapphire amplifier system that generates laser pulses with 1.55 eV photon energy and 90 fs pulse duration. The laser was split into two beams. One beam was frequency-converted through an optical parametric amplifier (OPA) and/or second harmonic generation for pumping the sample. The other beam generated THz pulses through a ZnTe crystal for probing the sample. By using the electro-optic sampling technique, we could map out the whole waveform of the THz electric field E(t) in the time domain, as well as the pump-induced change of the field $\Delta$E(t, $\tau$) [28, 29]. (Here t denotes the local time of the THz pulse and $\tau$ denotes the delay between the pump and probe pulse.)

We investigated large-area monolayer MoS$_2$ samples grown by chemical vapor deposition (CVD) [30, 31] on sapphire substrates. The electron doping density was estimated to be $n \approx 8 \pm 3 \times 10^{12}$ cm$^{-2}$ by transport measurements on similar samples. Detailed information on the fabrication and characterization of these CVD samples is provided in Ref. [31]. The samples were mounted on a helium-cooled cryostat in high vacuum. We first measured the THz absorption of monolayer MoS$_2$ samples in equilibrium conditions at T = 15 K. We recorded the THz electric field transmitted through the MoS$_2$/sapphire area, E(t), and as a reference, through an area without MoS$_2$ flakes, E$_0$(t) [Fig. 1(a)]. The maximum value of E(t) was 2.1% smaller than that of E$_0$(t). This attenuation of the THz transmission, as highlighted by the insets of Fig. 1(a), arises from the intraband absorption of excess free electrons in the doped sample. From the measured transmission spectra, we extracted the complex sheet conductivity of monolayer MoS$_2$ by applying the standard thin-film approximation [32-34]. We determined in the spectral range of 0.5 - 2.0 THz an average (real) conductivity $\sigma_1 \approx 2 \pm 0.5$ G$_0$, where G$_0$ = 2e$^2$/h is the quantum of conductance. This corresponds to an electron mobility $\mu = \sigma_1/ne \approx 120 \pm 50$ cm$^2$V$^{-1}$s$^{-1}$.

To investigate the optoelectronic response of monolayer MoS$_2$, we excited the sample using femtosecond laser pulses with photon energy 3.1 eV at T = 15 K. As this photon energy is sufficient to produce free electron-hole pairs in MoS$_2$, we expect an increase of THz absorption after photoexcitation. Surprisingly, we observed a significant increase of THz transmission after photoexcitation [Fig. 1(b)]. The pump-induced change of transmission (proportional to $\Delta$E) indicates a transient *decrease* of THz absorption in the monolayer MoS$_2$ sample [33]. We also measured the temporal ($\tau$) evolution of the fractional change of the THz field, that is, the ratio between the maximum values of waveforms $\Delta$E(t,$\tau$) and E(t) [Fig. 1(c)]. The THz dynamics exhibit a short component with lifetime $\tau_1 \approx 1$ ps, followed by a long component with lifetime $\tau_2 \approx 42$ ps. From the measured transmission spectra, we determined the corresponding change of complex sheet conductivity $\Delta\sigma(\omega,\tau)$ of monolayer MoS$_2$ with the thin-film approximation [33]. Both the real and imaginary parts of $\Delta\sigma(\omega,\tau)$ exhibit negative values for all delay times ($\tau$) and frequencies ($\omega$) in our measurement range [Fig. 1(d)].

The observed negative photoconductivity is a robust and substantial effect in monolayer MoS$_2$. It persists for all incident pump fluences (F = 0.4 – 170 $\mu$J/cm$^2$) and temperatures (T = 4 – 350 K) in our experiment [see, for example, fluence dependence in Fig. 2(a-b) and data for T = 300 K in Fig. 2(c)]. As the pump fluence increases, the reduction of the THz absorption increases markedly, gradually saturating at ~70% decrease of the total absorption [Fig. 2(b)], indicating that the conductivity of monolayer MoS$_2$ is reduced to only ~30% of its equilibrium value. The long component exhibit considerable increase of magnitude and relaxation time with the fluence, and become the dominant contribution to the overall dynamical response in the saturation regime [Fig. 2(a-b)]. This strong photo-reduction of conductivity is unexpected and counter-intuitive. For instance, an incident fluence F ~ 100 µJ/cm$^2$ in our experiment is expected to generate ~10$^{14}$ carriers/cm$^2$ in the MoS$_2$ monolayer [35], an order of magnitude higher than the doping density of our samples. One would expect a huge enhancement of THz absorption, instead of the strong bleaching observed in our experiment (e.g. see Fig. 11 in Ref. [27]).

Negative photoconductivity was observed in monolayer MoS$_2$ samples deposited on sapphire and quartz substrates, with and without HfO$_2$ or polymer electrolyte top layers. The insensitivity to the dielectric and interfacial environment implies that the phenomenon originates from the intrinsic property of the MoS$_2$ material. In addition, we examined multilayer CVD MoS$_2$ films and bulk MoS$_2$ crystals. We did not observe any photo-reduction of conductivity in these thicker samples, but only the expected photo-enhancement (see Supplemental Material [33]). The negative photoconductivity is therefore a unique phenomenon in the monolayer samples.

The substantially enhanced Coulomb interaction in the 2D limit is a distinct characteristic of monolayer MoS$_2$. The carrier interactions cause photo-generated electron-hole pairs to form bound neutral excitons that couple weakly with THz radiation. The binding energies of these excitons in MoS$_2$ and other TMD monolayers have been estimated to be a few hundred meV [13-22], an order of magnitude larger than the values of their respective multilayer and bulk crystals [15, 36]. In addition, when excess electrons are available, three-body bound states (two electrons and one hole) can be formed. These complex quasiparticles, the so-called trions or charged excitons, possess large dissociation energies (20 – 50 meV) and are stable even at room temperature [12, 16, 24, 25]. Trion formation can adequately account for our experimental observations: after pulsed excitation, the generated electron-hole pairs form trions with the excess free charges within the rise time of the pump-probe signal (~2 ps). The trions behave as free charged particles with increased effective mass [37], resulting in lower carrier mobility ($\mu$) and hence lower conductivity ($\sigma_1 = ne\mu$). In other words, instead of increasing the free-carrier concentration, photoexcitation adds weight to the original free charges and dulls their response to the electric field (Fig. 3).

A straightforward way to confirm the trion scheme is to compare the THz dynamics with the trion decay dynamics. We have carried out time-resolved PL measurements on our monolayer MoS$_2$ samples at T = 300 K [Fig. 2(d)]. The PL intensity in the emission photon energy range of 1.7 - 2.0 eV [inset of Fig. 5(a)], which is proportional to the trion population, decayed with a lifetime of $\tau_2 = 33 \pm 5$ ps. This decay time reflects the non-radiative recombination of trions in the defect sites, and is comparable to results in other experiments [38-41]. The trion lifetime agrees with the THz recovery time at similar conditions ($\tau_2 = 29$ ps) [Fig. 2(c)].

To further verify the trion mechanism, we have conducted the THz experiment with tunable pump photon energy from 0.77 to 2.11 eV (Fig. 4). The overall THz dynamical response is suppressed sharply and the major (long) component is quenched as the photon energy falls below 1.9 eV. This critical energy value matches the resonance of the A exciton and trion from the absorption spectrum of monolayer MoS$_2$ [Fig. 4(b)] [4, 24]. Near the excitonic resonance, where significant negative photoconductivity is observed, we expect that only excitons and trions are produced, with minimum free-carrier generation and carrier heating.

Trion formation provides a straightforward explanation for the ~70% reduction of THz conductivity in the saturation regime [Fig. 2(b)], which simply reflects the three-fold increase of carrier effective mass. In this picture, the photoconductivity saturates at a value proportional to the doping electron density, when a large fraction of these excess electrons join excitons to form trions. This is consistent with the results in Fig. 2(b), where the excitation density (~$10^{13}$ cm$^{-2}$) at the onset of saturation (F ~10 μJ/cm$^2$) matches roughly the doping density of our sample (~8 x $10^{12}$ cm$^{-2}$). To further explore this behavior, we have examined the THz dynamics and PL of an MoS$_2$ sample at different doping stages. At the initial stage, the pristine or as-grown sample exhibited pronounced THz response (black line in Fig. 5). Its PL spectrum was centered at the trion (A$^-$) recombination energy (1.844 eV), indicating strong unintentional electron doping from the substrate [inset of Fig. 5(a)]. We next doped the sample with chemical solutions according to the method described in Ref. [42]. We first deposited F$_4$TCNQ molecules as *p*-type chemical dopants to lower the electron density of the sample [42]. The PL of the sample was found to increase and the peak position blueshifted to the exciton energy (1.885 eV), indicating the suppression of trion formation [24, 42]. Correspondingly, the THz response was suppressed and subsequently quenched. The response also saturated at lower pump fluence (red and green lines in Fig. 5). The process could be reversed by depositing NADH molecules as *n*-type dopants to increase the electron density [42]. Both the PL spectrum and the THz response were partially recovered (blue lines in Fig. 5). The observed doping dependence agrees qualitatively with the expected behavior of trion formation.

Furthermore, the trion scheme naturally accounts for the fluence and photon-energy dependence of the THz recovery time. As shown in the insets of Fig. 2(a) and 4(a), the long-component lifetime ($\tau_2$) increases markedly with increasing pump fluence or photon energy, both corresponding to the increase of trion density. The excitons and trions in MoS$_2$ are known to decay predominantly through defect-mediated processes. As their density increases, the defect sites will be gradually filled up and the probability to trap a trion decreases accordingly. This will thus prolong the trion lifetime and hence, the THz recovery time.

The evidence and analysis presented above strongly suggest that trion formation is the dominant mechanism for the negative photoconductivity in monolayer MoS$_2$, particularly the major (long) component of the THz dynamics. We now consider other secondary factors, which could become important at shorter time scales. The short component of the THz dynamics increases with pump fluence and photon energy, with only a slight change of lifetime (1-3 ps) [see the difference between the blue and red dots in Fig. 2(b) and 4(b)]. It also exhibits a step-like increase as the pump photon energy reaches the trion generation energy (~1.85 eV) [Fig. 4(b)]. These findings suggest that trion-trion annihilation might be important at excitation energies above 1.85 eV. A significant fraction of the short component is, however, retained even when the pump photon energy drops below 1.85 eV, and persists down to 0.77 eV [Fig. 4]. Such sub-gap response implies non-negligible optical absorption by free carriers through defect-mediated processes. The resultant transient increase of carrier temperature might reduce the mobility, and hence the conductivity, of monolayer MoS$_2$. The hot carriers in MoS$_2$ are known to relax within ~1 ps by emission of phonons [43], and the optical phonons decay in ~1 ps as estimated from their Raman line widths (4 – 8 cm$^{-1}$) [31]. Both time scales match the short-component lifetime of the THz dynamics. Similar ultrafast reduction of THz conductivity by pulsed laser heating with ~2 ps recovery time has been observed in graphene, a semi-metallic 2D material with no trions [28, 29, 34, 44]. We therefore tentatively attribute the short component under sub-gap optical excitation to the laser heating effect. More investigations are needed to elucidate the roles of these different mechanisms in the photoconductivity behaviors of monolayer MoS$_2$.

In conclusion, we have observed a dramatic reduction of THz conductivity in monolayer MoS$_2$ under optical excitation. This unusual phenomenon originates from the strong many-body interactions in the material, which convert the 2D electron gas into a charged trion gas with the same charge density but much heavier effective mass. Similar trionic phenomena should prevail in other monolayer TMDs. The interaction-driven modulation of conductivity enriches the physical properties of 2D materials and provides new concepts in developing novel optoelectronic and excitonic devices.

We thank K. F. Mak, T. F. Heinz, X. D. Xu, W. R. Lambrecht, and V. Bulović for helpful discussions, L. Yu and E. J. Sie for transport and optical characterization of the samples. This research was supported by Department of Energy Office of Basic Energy Sciences Grant No DE-SC0006423. Y.H.L and J.K. acknowledge support from the National Science Foundation under award number NSF DMR 0845358.

**References:**

[1] G. D. Scholes, and G. Rumbles, Nature Mater. **5**, 683-696 (2006).
[2] Q. H. Wang, K. Kalantar-Zadeh, A. Kis, J. N. Coleman, and M. S. Strano, Nature Nanotech. **7**, 699-712 (2012).
[3] S. Z. Butler, S. M. Hollen, L. Cao, Y. Cui, J. A. Gupta, H. R. Gutiérrez, T. F. Heinz, S. S. Hong, J. Huang, A. F. Ismach, E. Johnston-Halperin, M. Kuno, V. V. Plashnitsa, R. D. Robinson, R. S. Ruoff, S. Salahuddin, J. Shan, L. Shi, M. G. Spencer, M. Terrones, W. Windl, and J. E. Goldberger, ACS Nano **7**, 2898-2926 (2013).
[4] K. F. Mak, C. Lee, J. Hone, J. Shan, and T. F. Heinz, Phys. Rev. Lett. **105**, 4 (2010).
[5] A. Splendiani, L. Sun, Y. Zhang, T. Li, J. Kim, C.-Y. Chim, G. Galli, and F. Wang, Nano Lett. **10**, 1271-1275 (2010).
[6] B. Radisavljevic, A. Radenovic, J. Brivio, V. Giacometti, and A. Kis, Nat. Nanotech. **6**, 147-150 (2011).
[7] O. Lopez-Sanchez, D. Lembke, M. Kayci, A. Radenovic, and A. Kis, Nat. Nanotech. **8**, 497-501 (2013).
[8] K. F. Mak, K. L. He, J. Shan, and T. F. Heinz, Nat. Nanotech. **7**, 494-498 (2012).
[9] H. L. Zeng, J. F. Dai, W. Yao, D. Xiao, and X. D. Cui, Nat. Nanotech. **7**, 490-493 (2012).
[10] T. Cao, G. Wang, W. P. Han, H. Q. Ye, C. R. Zhu, J. R. Shi, Q. Niu, P. H. Tan, E. Wang, B. L. Liu, and J. Feng, Nat. Commun. **3**, 5 (2012).
[11] X. Xu, W. Yao, D. Xiao, and T. F. Heinz, Nat. Phys. **10**, 343-350 (2014).
[12] A. M. Jones, H. Y. Yu, N. J. Ghimire, S. F. Wu, G. Aivazian, J. S. Ross, B. Zhao, J. Q. Yan, D. G. Mandrus, D. Xiao, W. Yao, and X. D. Xu, Nat. Nanotech. **8**, 634-638 (2013).
[13] T. Cheiwchanchamnangij, and W. R. L. Lambrecht, Phys. Rev. B **85**, 205302 (2012).
[14] A. Ramasubramaniam, Phys. Rev. B **86**, 115409 (2012).
[15] H.-P. Komsa, and A. V. Krasheninnikov, Phys. Rev. B **86**, 241201 (2012).
[16] T. C. Berkelbach, M. S. Hybertsen, and D. R. Reichman, Phys. Rev. B **88**, 045318 (2013).
[17] D. Y. Qiu, F. H. da Jornada, and S. G. Louie, Phys. Rev. Lett. **111**, 216805 (2013).
[18] A. Chernikov, T. C. Berkelbach, H. M. Hill, A. Rigosi, Y. Li, O. B. Aslan, D. R. Reichman, M. S. Hybertsen, and T. F. Heinz, arXiv:1403.4270 (2014).
[19] Z. Ye, T. Cao, K. O'Brien, H. Zhu, X. Yin, Y. Wang, S. G. Louie, and X. Zhang, arXiv:1403.5568 (2014).
[20] B. Zhu, X. Chen, and X. Cui, arXiv:1403.5108 (2014).

[21] M. M. Ugeda, A. J. Bradley, S.-F. Shi, F. H. da Jornada, Y. Zhang, D. Y. Qiu, S.-K. Mo, Z. Hussain, Z.-X. Shen, F. Wang, S. G. Louie, and M. F. Crommie, arXiv:1404.2331 (2014).
[22] G. Wang, X. Marie, I. Gerber, T. Amand, D. Lagarde, L. Bouet, M. Vidal, A. Balocchi, and B. Urbaszek, arXiv:1404.0056 (2014).
[23] E. J. Sie, Y.-H. Lee, A. J. Frenzel, J. Kong, and N. Gedik, arXiv:1312.2918 (2014).
[24] K. F. Mak, K. L. He, C. Lee, G. H. Lee, J. Hone, T. F. Heinz, and J. Shan, Nat. Mater. **12**, 207-211 (2013).
[25] J. S. Ross, S. F. Wu, H. Y. Yu, N. J. Ghimire, A. M. Jones, G. Aivazian, J. Q. Yan, D. G. Mandrus, D. Xiao, W. Yao, and X. D. Xu, Nat. Commun. **4**, 6 (2013).
[26] R. Ulbricht, E. Hendry, J. Shan, T. F. Heinz, and M. Bonn, Rev. Mod. Phys. **83**, 543-586 (2011).
[27] M. C. Beard, G. M. Turner, and C. A. Schmuttenmaer, Phys. Rev. B **62**, 15764 (2000).
[28] A. J. Frenzel, C. H. Lui, W. Fang, N. L. Nair, P. K. Herring, P. Jarillo-Herrero, J. Kong, and N. Gedik, App. Phys. Lett. **102**, - (2013).
[29] A. J. Frenzel, C. H. Lui, Y. C. Shin, J. Kong, and N. Gedik, arXiv:1403.3669 (2014).
[30] Y.-H. Lee, X.-Q. Zhang, W. Zhang, M.-T. Chang, C.-T. Lin, K.-D. Chang, Y.-C. Yu, J. T.-W. Wang, C.-S. Chang, L.-J. Li, and T.-W. Lin, Adv. Mater. **24**, 2320-2325 (2012).
[31] Y.-H. Lee, L. Yu, H. Wang, W. Fang, X. Ling, Y. Shi, C.-T. Lin, J.-K. Huang, M.-T. Chang, C.-S. Chang, M. Dresselhaus, T. Palacios, L.-J. Li, and J. Kong, Nano Lett. **13**, 1852-1857 (2013).
[32] M. C. Nuss, and J. Orenstein, in *Millimeter and Submillimeter Wave Spectroscopy of Solids,* G. Grüner, Ed. (1998).
[33] See Supplemental Material for detailed information of experiment and analysis, and THz study of multilayer $MoS_2$.
[34] G. Jnawali, Y. Rao, H. Yan, and T. F. Heinz, Nano Lett. **13**, 524-530 (2013).
[35] K. F. Mak, C. Lee, J. Hone, J. Shan, and T. F. Heinz, Phys. Rev. Lett. **105**, 136805 (2010).
[36] E. Fortin, and F. Raga, Phys. Rev. B **11**, 905-912 (1975).
[37] D. Sanvitto, F. Pulizzi, A. J. Shields, P. C. M. Christianen, S. N. Holmes, M. Y. Simmons, D. A. Ritchie, J. C. Maan, and M. Pepper, Science **294**, 837-839 (2001).
[38] T. Korn, S. Heydrich, M. Hirmer, J. Schmutzler, and C. Schuller, Appl. Phys. Lett. **99**, 102109-102103 (2011).
[39] H. Shi, R. Yan, S. Bertolazzi, J. Brivio, B. Gao, A. Kis, D. Jena, H. G. Xing, and L. Huang, ACS Nano **7**, 1072-1080 (2012).
[40] R. Wang, B. A. Ruzicka, N. Kumar, M. Z. Bellus, H.-Y. Chiu, and H. Zhao, Phys. Rev. B **86**, 045406 (2012).
[41] A. M. van der Zande, P. Y. Huang, D. A. Chenet, T. C. Berkelbach, Y. You, G.-H. Lee, T. F. Heinz, D. R. Reichman, D. A. Muller, and J. C. Hone, Nature Mater. **12**, 554–561 (2013).
[42] S. Mouri, Y. Miyauchi, and K. Matsuda, Nano Lett. **13**, 5944-5948 (2013).
[43] A. Tanaka, N. J. Watkins, and Y. Gao, Phys. Rev. B **67**, 113315 (2003).
[44] S. F. Shi, T. T. Tang, B. Zeng, L. Ju, Q. Zhou, A. Zettl, and F. Wang, Nano Lett. **14**, 1578-1582 (2014).

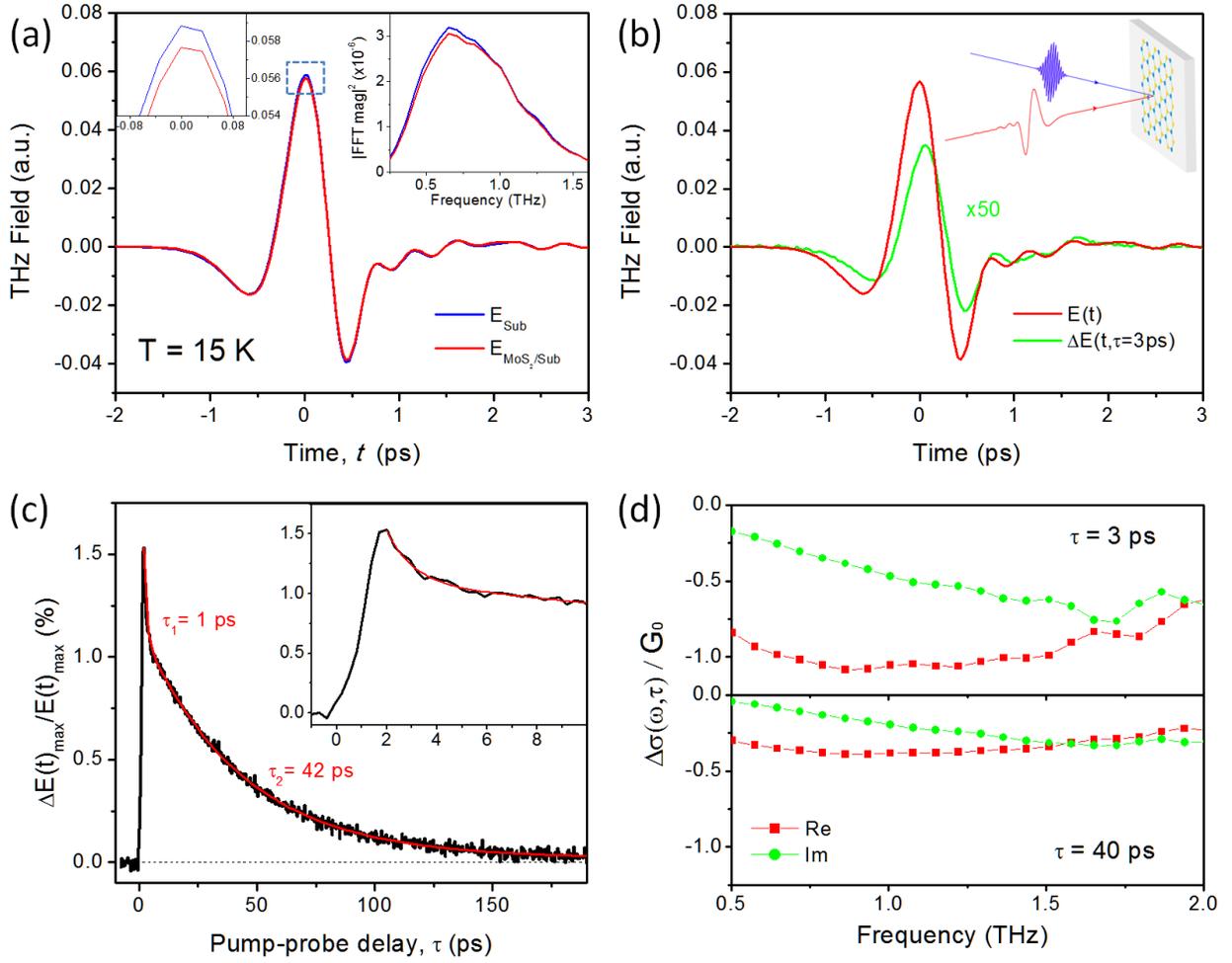

FIG. 1. (a) THz response of monolayer $MoS_2$ on a sapphire substrate in equilibrium conditions. The blue and red lines show the THz electric field transmitted through areas with and without the sample, respectively. The left inset is a zoomed-in view of the peak (dashed square). The measurement uncertainty at the peak is $3.5 \times 10^{-5}$, corresponding to 0.06% of the total THz signal, much smaller than the 2.1% attenuation of THz field by the $MoS_2$ sample. The right inset shows the power spectra of the corresponding time-domain signals. (b) Change of THz transmission after pulsed excitation. The red line denotes the equilibrium THz field E(t) transmitted through the $MoS_2$ sample. The green line denotes the pump-induced waveform, $\Delta E(t, \tau)$, scaled by a factor of 50, at pump-probe delay $\tau = 3$ ps. The inset shows a schematic of our optical-pump THz-probe experiment on monolayer $MoS_2$. (c) Temporal evolution of the ratio between the maximum values of waveforms $\Delta E(t, \tau)$ and $E(t)$. The red line is a biexponential fit, yielding lifetimes $\tau_1 = 1$ ps and $\tau_2 = 42$ ps. The inset is a zoomed-in view of the fast decay component. (d) The extracted pump-induced change of complex sheet conductivity $\Delta\sigma(\omega,\tau)$ of monolayer $MoS_2$ at $\tau = 3$ and 40 ps. All measurements were made at T = 15 K. The excitation photon energy was 3.1 eV and the incident pump fluence was 50 μJ/cm².

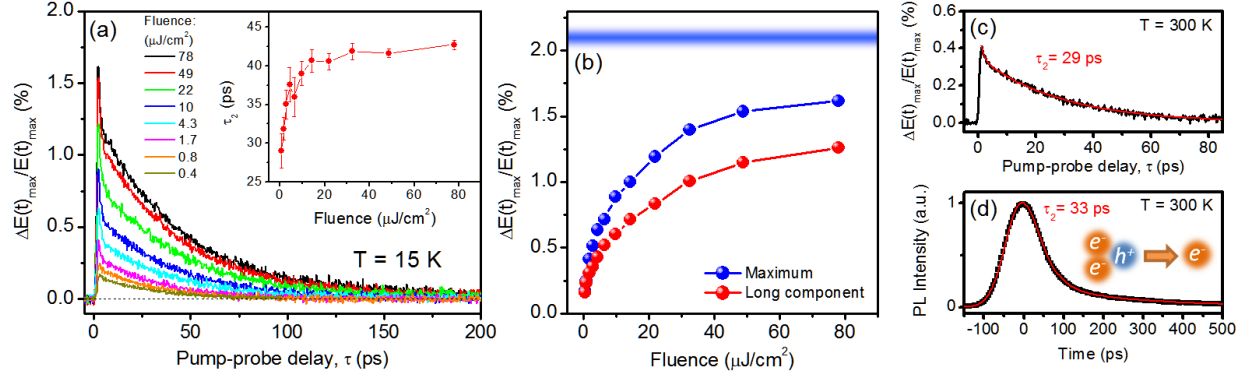

FIG. 2 (a) Temporal THz dynamics of monolayer MoS$_2$ at T = 15 K, as in Fig. 1(c), for different incident pump fluences, with excitation photon energy 3.1 eV. All spectra show a two-component decay process. The inset shows the long-component lifetime ($\tau_2$), extracted from biexponential fits, as a function of fluence. No systematic fluence dependence was found for the short-component lifetime ($\tau_1 \approx 1$ ps). (b) The maximum pump-induced signal (at $\tau \approx 2$ ps) and the magnitude of the long component as a function of fluence. The blue bar at 2.1 ± 0.06 % denotes the total attenuation of THz field by the unexcited MoS$_2$ sample, as determined from Fig. 1(a). (c) THz dynamics at room temperature. A biexponential fit (red line) yields lifetimes $\tau_1 = 1.1$ ps and $\tau_2 = 29$ ps. (d) Time-resolved photoluminescence (PL) of monolayer MoS$_2$ at emission photon energies 1.7 - 2.0 eV measured by time-correlated single photon counting. The red line is a biexponential fit, convolved with a Gaussian instrumental response function (standard deviation 36 ps). The best-fit lifetimes are $\tau_2 = 33 \pm 5$ ps and $\tau_3 = 230 \pm 10$ ps. The inset depicts trion recombination as the underlying mechanism for the PL dynamics. In (c) and (d), we used pump fluence ~200 µJ/cm$^2$ and excitation photon energy 3.1 eV.

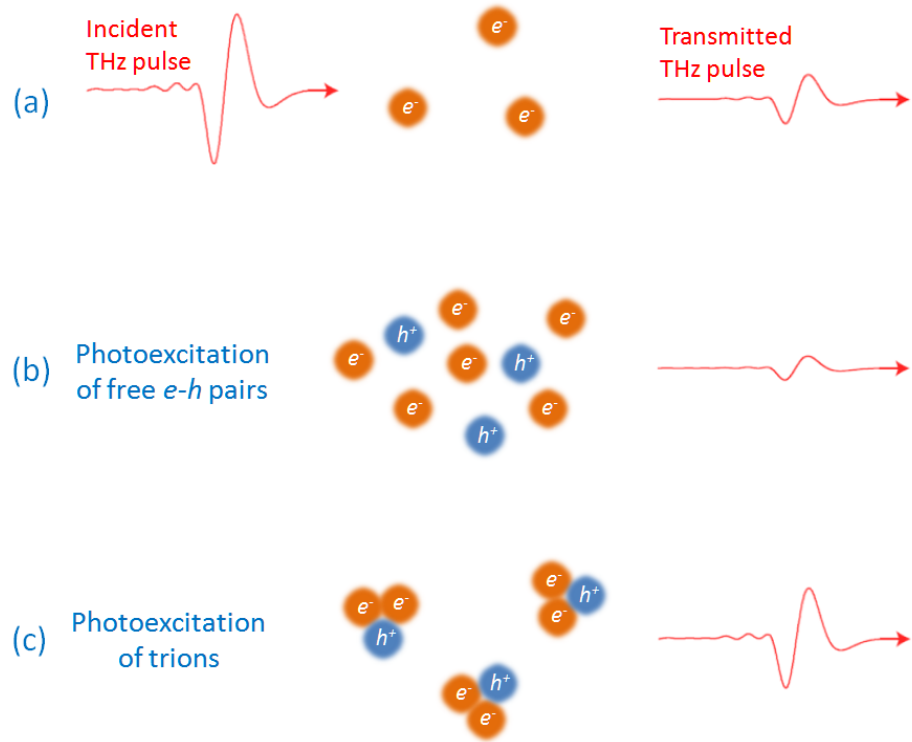

FIG. 3. Schematic of trionic effect on the THz response of a semiconductor under interband photoexcitation. (a) Absorption of THz pulse by doping-induced free carriers. (b) Increased THz absorption by excitation of free electron-hole pairs. (c) Reduced THz absorption by formation of trions.

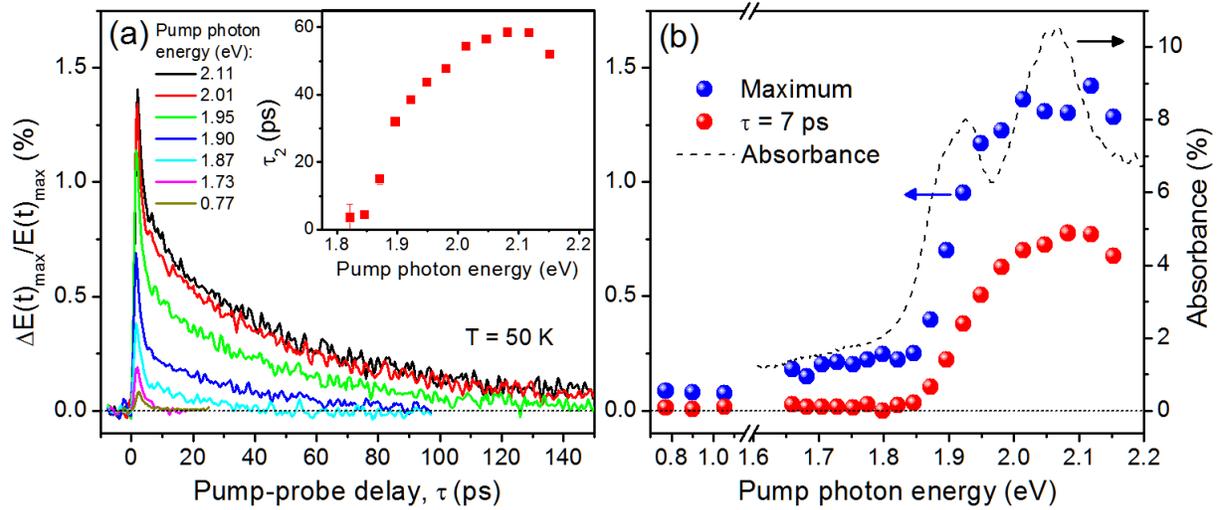

FIG. 4. (a) Temporal THz dynamics of monolayer $MoS_2$ at T = 50 K, with the same incident pump fluence (65 µJ/cm$^2$) but different pump photon energy (0.77 – 2.11 eV). The inset shows the long-component lifetime ($\tau_2$), extracted from biexponential fits, as a function of excitation photon energy. No systematic energy dependence was found for the short-component lifetime ($\tau_1 \approx 2$ ps). (b) The maximum pump-induced signal at $\tau \approx 2$ ps and at $\tau = 8$ ps (representing the long-component contribution) as a function of excitation photon energy. The dashed curve is the absorption spectrum measured on a representative monolayer $MoS_2$ flake on sapphire at T = 20 K according to the method described in Ref. [4].

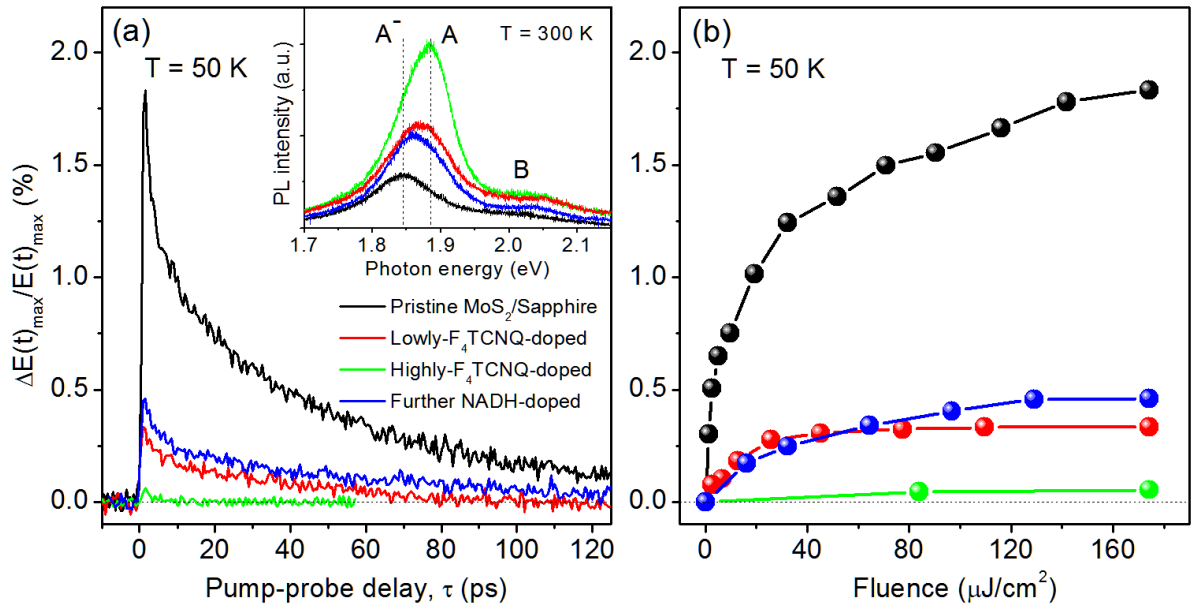

FIG. 5. (a) Temporal THz dynamics of monolayer $MoS_2$ at different chemical doping stages. The measurements were made on a pristine $MoS_2$ sample on sapphire (black), the same sample after being doped lightly (red) and heavily (green) with p-type $F_4TCNQ$ molecules and after being further doped with n-type NADH molecules. The sample is different from that used in Fig. 2. All measurements were made with the same pump photon energy (1.98 eV) and fluence (174 $\mu J/cm^2$) at T = 50 K. The inset shows the photoluminescence (PL) spectra of the $MoS_2$ sample at the corresponding doping stages, taken at room temperature with 532-nm laser excitation. The position of A and B excitons and trion ($A^-$) is denoted. (b) The maximum pump-induced signal (at $\tau \approx 2$ ps) as a function of pump fluence at the corresponding doping stage at (a).

# *Supplemental Material*

## 1. Experimental Methods

We measured the low-frequency optical response of monolayer $MoS_2$ by using time-domain THz spectroscopy in conjunction with optical pump excitation. The laser source is a Ti:sapphire regenerative amplifier system (Spitfire Pro, Spectra-Physics) that generates pulses of central wavelength $\lambda = 800$ nm (1.55eV), pulse duration ~90 fs, at a repetition rate 5 kHz. The 800-nm laser was split into two beams. One beam was frequency-converted for pumping the samples. The photon energy of the pump beam can be doubled to 3.1 eV by second harmonic generation in a BBO crystal, or converted to the infrared range (0.77 – 1.1 eV) by using an optical parametric amplifier (OPA) (TOPAS-Prime, Spectra-Physics). The photon energy of the infrared pump beam can further be doubled to 1.6 – 2.2 eV by second harmonic generation. The other beam was used to generate THz pulses through a 1-mm-thick ZnTe crystal for probing the samples. The pump and probe pulses were synchronized with a controllable delay ($\tau$) and irradiated the sample with spot diameter of ~6 mm and ~3 mm, respectively. The samples were mounted on a helium-cooled cryostat in high vacuum (pressure < $10^{-5}$ Torr). The THz beam transmitted through the sample was detected by electro-optic sampling with the 800-nm pulses through another 1-mm-thick ZnTe crystal. We made use of a data acquisition (DAQ) card synchronized with a phase-locked mechanical chopper to collect the signal for each pulse in the experiment [S1]. We chopped the THz probe beam to measure the THz field profile E(t) in time domain, and chopped the optical pump beam to measure the pump-induced change of THz field $\Delta E(t, \tau)$.

The time-resolved photoluminescence (PL) measurement were conducted by time-correlated single photon counting at room temperature in ambient conditions. The CVD monolayer $MoS_2$ samples were excited with a frequency-doubled Ti:Sapphire laser at $\lambda = 400$ nm (3.1 eV) with pulse duration ~100 fs and repetition rate 40 MHz. The laser was focused onto the sample with a spot diameter ~3 μm through a 20× microscope objective. The PL was collected by the same objective, passed through a dichroic beam-splitter and a longpass filter ($\lambda = 600$ nm), and directed onto a single photon counting avalanche photodiode. The total instrument response function was ~36 ps, as characterized by the rise time of the PL signal. Although the measurement detected PL from both neutral excitons and charged trions, most of the PL signal was from the trions in our doped samples (see the PL spectrum at the inset of Fig. 5(a) in the main paper) [S2].

## 2. Extraction of conductivity

To determine the THz conductivity of monolayer $MoS_2$ samples in equilibrium, we have recorded the THz field transmitted through the $MoS_2$/sapphire area, E(t), and as a reference, through the sapphire area without $MoS_2$ flakes, $E_0(t)$. Fourier transformation of the field waveform yields the corresponding (complex) frequency-domain fields, $E(\omega)$ and $E_0(\omega)$. Applying the standard thin-film approximation, we extract the complex sheet conductivity $\sigma(\omega)$ of the $MoS_2$ monolayer using the following formula [S3, 4]

$$\frac{E(\omega)}{E_0(\omega)} = \frac{n_s + 1}{n_s + 1 + Z_0 \sigma(\omega)} \tag{S1}$$

Here $n_s = 3.07$ is the refractive index of the sapphire substrate at ~1 THz and $Z_0 = 377$ Ω is the impedance of free space. We have also recorded the pump-induced change of THz field $\Delta E(t)$, and compared it with the total transmitted field E(t). Using the differential form of Eq. (S1), we obtain the differential optical conductivity $\Delta\sigma(\omega)$ as [S3-5]:

$$\Delta\sigma(\omega) = -\left(\frac{n_s+1}{Z_0}\right)\frac{\Delta E(\omega)}{E(\omega)} \quad (S2)$$

This relation shows that an increase of transmission corresponds to a decrease of THz conductivity in our experimental geometry.

### 3. Time-resolved THz spectroscopy on multilayer MoS$_2$ samples

Beside the study of monolayer MoS$_2$ presented in the main paper, we have also carried out comparative investigations on multilayer MoS$_2$ systems. The multilayer MoS$_2$ samples were grown by chemical vapor deposition (CVD) on sapphire substrates with a prolonged growing time as compared to the monolayer samples [S6]. The sample thickness is estimated to be 5-10 layers by using optical contrast in the visible range.

In the optical-pump THz-probe measurement of multilayer MoS$_2$, we observed a transient decrease of transmitted THz field after optical excitation with incident pump fluence F ≈ 80 μJ/cm$^2$ and excitation photon energy 3.1 eV at T = 15 K [Fig. S1(a)]. This corresponds to enhanced absorption in the multilayer MoS$_2$ samples, in contrast to the reduced absorption observed in monolayer [Fig. 1(b) in the main paper]. We have also measured the temporal evolution of the fractional change of the THz electric field $\Delta E(t,\tau)$ at different pump-probe delay time. Here t denotes the local time of the THz pulse and $\tau$ denotes the delay between the pump and probe pulse. In this case of reduced transmission, the change of THz electric field is represented by the ratio between the minimum value of waveform $\Delta E(t,\tau)$ and the maximum value of waveform E(t) [Fig. S1(b)]. The THz dynamics exhibited two decay components with lifetimes $\tau_1$ = 1.7 ps and $\tau_2$ = 18 ps. We also observed a long component that persisted for over 200 ps. From the measured $\Delta E(t,\tau)$ profile, we can determine the corresponding pump-induced change of conductivity $\Delta\sigma(\omega,\tau) = \Delta\sigma_1(\omega,\tau) + i\Delta\sigma_2(\omega,\tau)$ using Eq. (S2). Fig. S1(c) displays the $\Delta\sigma(\omega,\tau)$ spectra at pump-probe delay $\tau$ = 4.5 ps. Both the real and imaginary parts of $\Delta\sigma(\omega,\tau)$ are positive for all frequencies in our measured frequency range. The result stands in contrast to the negative differential conductivity observed in monolayers and is similar to those of conventional semiconductors with weak excitonic effects [S7, 8].

The $\Delta\sigma(\omega,\tau)$ spectra in our measurement can be described roughly by the Drude formula,

$$\Delta\sigma(\omega) = \frac{Ne^2}{m}\frac{1}{\gamma - i\omega}. \quad (S3)$$

Here N, $\gamma$, e, m denote the photo-generated free carrier density, carrier scattering rate, electronic charge, and carrier effective mass, respectively. For our n-doped multilayer MoS$_2$ sample, we consider the electrons around the conduction minimum along the Γ-K direction in the Brillouin Zone, which have effective mass $m \approx 0.6\, m_e$, where $m_e$ is the electron rest mass. By fitting our data with Eq. (S3), we extract N = 4 × 10$^{11}$ cm$^{-2}$ and $\gamma$ = 7.2 THz [Fig S1(c)]. From these parameters, we estimate the mobility to be $\mu = e/m\gamma \approx$ 400 cm$^2$/Vs for our multilayer CVD MoS$_2$ samples at T = 15 K.

We note that the extracted charge density N = 4 × 10$^{11}$ cm$^{-2}$ is much smaller than the value (~10$^{14}$ cm$^{-2}$) expected from the incident pump fluence F = 80 μJ/cm$^2$. This suggests that the excitonic effects are still significant in multilayer MoS$_2$ samples, though they are weaker than those in monolayer MoS$_2$. Most photo-excited electron-hole pairs therefore still form neutral excitons that interact weakly with THz radiation. This scenario is supported by the still sizable exciton binding energies reported in bulk MoS$_2$ (~80 meV) [S9, 10]. The enhanced conductivity observed in our experiment should therefore arise from a small fraction of residual free carriers. In this scenario, we tentatively interpret the fast decay of the positive differential conductivity as the condensation of these residual free charges into excitons. As we did not observe any reduction of conductivity, the trionic effect should be unimportant in the multilayer system.

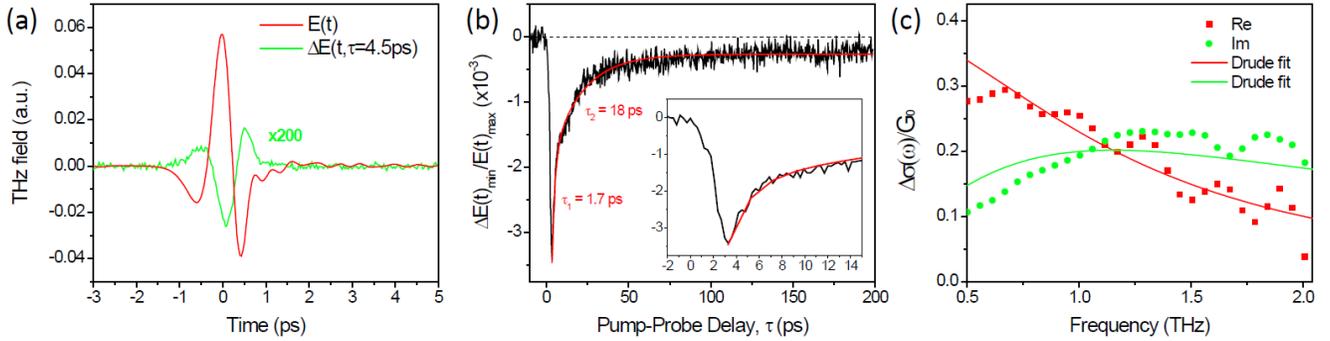

FIG. S1. Photoinduced increase of THz conductivity in CVD multilayer MoS$_2$ samples at T = 15 K. (a) Change of THz transmission after pulsed optical excitation. The red line denotes the THz electric field E(t) transmitted through the MoS$_2$ sample in equilibrium. The green line denotes the pump-induced waveform, ΔE(t, τ), scaled by a factor of 200, at a pump-probe delay τ = 4.5 ps. (b) Temporal evolution of the fractional change of the transmitted THz field after pulsed excitation. The y-axis represents the ratio between the minimum value of waveform ΔE(t, τ) and the maximum value of waveform E(t). The red line is a biexponential fit of the data, yielding lifetimes $\tau_1$ = 1.7 ps and $\tau_2$ = 18 ps. A slower component exceeding 200 ps is also observed in the data. The inset is a zoomed-in view of the fast decay component. (c) The extracted pump-induced change of the complex sheet conductivity Δσ(ω,τ) of multilayer MoS$_2$ at pump-probe delay τ = 4.5 ps. The red squares and green dots represent, respectively, the real and imaginary part of Δσ(ω,τ). The lines are fits by a Drude model [Eq. (S3)]. In all measurements, the photon energy, pulse duration and incident pump fluence of the excitation laser were 3.1 eV, 100 fs and 80 μJ/cm$^2$, respectively.

**Supplemental references:**

[S1] C. A. Werley, S. M. Teo, and K. A. Nelson, Review of Scientific Instruments **82**, 123108-123106 (2011).
[S2] K. F. Mak, K. He, C. Lee, G. H. Lee, J. Hone, T. F. Heinz, and J. Shan, Nature Mater. **12**, 207-211 (2013).
[S3] M. C. Nuss, and J. Orenstein, in *Millimeter and Submillimeter Wave Spectroscopy of Solids,* G. Grüner, Ed. (1998).
[S4] G. Jnawali, Y. Rao, H. Yan, and T. F. Heinz, Nano Lett. **13**, 524-530 (2013).
[S5] A. J. Frenzel, C. H. Lui, W. Fang, N. L. Nair, P. K. Herring, P. Jarillo-Herrero, J. Kong, and N. Gedik, Appl. Phys. Lett. **102**, 113111 (2013).
[S6] Y.-H. Lee, L. Yu, H. Wang, W. Fang, X. Ling, Y. Shi, C.-T. Lin, J.-K. Huang, M.-T. Chang, C.-S. Chang, M. Dresselhaus, T. Palacios, L.-J. Li, and J. Kong, Nano Lett. **13**, 1852-1857 (2013).
[S7] M. C. Beard, G. M. Turner, and C. A. Schmuttenmaer, Phys. Rev. B **62**, 15764 (2000).
[S8] R. Ulbricht, E. Hendry, J. Shan, T. F. Heinz, and M. Bonn, Rev. Mod. Phys. **83**, 543-586 (2011).
[S9] E. Fortin, and F. Raga, Phys. Rev. B **11**, 905-912 (1975).
[S10] H.-P. Komsa, and A. V. Krasheninnikov, Phys. Rev. B **86**, 241201 (2012).